\documentstyle[12pt]{article}
\def\MeV{\,{\rm MeV}}

\def\cmm2{{\,\rm cm^{-2}}}
\def\cm2{{\,{\rm cm}^2}}
\def\cmm3{{\,{\rm cm}^{-3}}}
\def\gcmm3{{\,{\rm g\,cm^{-3}}}}

\def\la{\mathrel{\mathpalette\fun <}}
\def\ga{\mathrel{\mathpalette\fun >}}
\def\fun#1#2{\lower3.6pt\vbox{\baselineskip0pt\lineskip.9pt
  \ialign{$\mathsurround=0pt#1\hfil##\hfil$\crcr#2\crcr\sim\crcr}}}

\def\beginapjbib{\begingroup \section*{References}
         \parskip=.5ex plus 1.0pt
	 \def\bibitem{\par \noindent \hangindent\parindent
		\hangafter=1}}
\def\endapjbib{\par \endgroup}

\begin{document}
\pagestyle{empty}
\begin{center}
\bigskip

\null
\vskip 1.8in
\vspace{.2in}
{\Large \bf Big-Bang Nucleosynthesis\\
\bigskip
and Galactic Chemical Evolution}
\bigskip

\vspace{.2in}
Craig J.~Copi,$^{1,2}$ David N.~Schramm,$^{1,2,3}$ and
Michael S. Turner$^{1,2,3}$\\

\vspace{.2in}
{\it $^1$Department of Physics \\
Enrico Fermi Institute, The University of Chicago, Chicago, IL~~60637-1433}\\

\vspace{0.1in}
{\it $^2$NASA/Fermilab Astrophysics Center\\
Fermi National Accelerator Laboratory, Batavia, IL~~60510-0500}\\

\vspace{0.1in}
{\it $^3$Department of Astronomy \& Astrophysics\\
The University of Chicago, Chicago, IL~~60637-1433}\\

\end{center}

\vspace{.3in}
\centerline{\bf ABSTRACT}
\bigskip

Deuterium is the best indicator of the baryon density; however,
only its present abundance is known (and only locally)
and its chemical evolution is intertwined with that of $^3$He.
Because galactic abundances are spatially heterogeneous,
mean chemical-evolution models are not well suited for extrapolating
the pre-solar D and $^3$He abundances to their primeval
values.  We introduce a new approach which explicitly addresses heterogeneity,
and show that the decade-old big-bang nucleosynthesis concordance
interval $\eta \approx (2 -8)\times 10^{-10}$
based on D and $^3$He is robust.

\vfill
submitted to {\it Astrophysical Journal Letters}

\newpage
\pagestyle{plain}
\setcounter{page}{1}
\newpage
\section{Introduction}

Big-bang nucleosynthesis is the earliest and most stringent test of the
standard cosmology.  The inferred primeval abundances of
D, $^3$He, $^4$He and $^7$Li are in accord with the
big-bang predictions provided that the baryon-to-photon ratio $\eta$
between about $2.5\times 10^{-10}$ and $6\times 10^{-10}$, which
corresponds to $\Omega_B \simeq 0.009h^{-2} - 0.02h^{-2}$,
and the number of light (mass less than about $1\MeV$) particle species,
expressed as the equivalent number of massless
neutrino species, $N_\nu \le 3.4$
(Walker et al.~1991; Krauss \& Kernan~1995; Copi, Schramm, \& Turner 1995).
Big-bang nucleosynthesis thereby provides the best determination
of the density of ordinary matter and an important constraint to
theories that attempt to unify the fundamental forces and particles of Nature.

Of the light elements D has the most potential as a ``baryometer''
because its production depends sensitively upon $\eta$,
D/H $\propto \eta^{-1.7}$.  On the other hand, its
interpretation is challenging because D is burned in
virtually all astrophysical situations and its abundance has only been
measured in the solar vicinity.  Because D is destroyed and not
produced (Epstein, Lattimer, \& Schramm~1976) a firm upper limit to $\eta$
can be obtained by insisting that big-bang production accounts for the D
observed in the local ISM, D/H $\ge (1.6\pm 0.1)
\times 10^{-5}$ (Linsky et al.~1993).
This leads to the two-decade old bound, $\eta\la 9\times 10^{-10}$, which is
the linchpin in the argument that baryons cannot provide closure density
(Reeves et al.~1973).

Because D is so readily destroyed, it is not possible to obtain a
lower bound to $\eta$ based upon D alone.
The sum of D and $^3$He is more promising:
Since deuterium is first burned to $^3$He, and $^3$He is much more
difficult to burn, (D + $^3$He)/H is much less sensitive to chemical evolution
(Yang et al.~1984).  This argument depends upon the
fraction of $^3$He that survives stellar processing (referred to as $g_3$),
which itself depends strongly upon stellar mass: high mass stars destroy
$^3$He, $g_3 \sim 0.2$, whereas low-mass stars are believed to
produce additional $^3$He
(Iben and Truran~1978; Dearborn, Schramm, Steigman~1986).

To obtain a lower bound to $\eta$, one needs good determinations of both
D and $^3$He in the same place, at the same time, as well as a lower
bound to $g_3$.  The only place where both abundances are known
is the pre-solar nebula, (D/H)$_\odot\sim 2.7
\times 10^{-5}$ and $(^3$He/H)$_\odot \sim 1.5 \times 10^{-5}$.
Arguing that the mean survival fraction ${\bar g}_3$ is
greater than 0.25, the lower limit, $\eta \ga 2.5\times 10^{-10}$
has been derived (Yang et al.~1984).  Thus, D and $^3$He together
define a big-bang consistency interval,
$\eta \simeq (2.5 - 9) \times 10^{-10}$.

Our aim here is to provide a firmer basis for these arguments.  To
this end we introduce a new approach to the chemical evolution of D and
$^3$He which allows their primeval abundances to be extracted
from pre-solar abundances, while explicitly including heterogeneity.
By considering an extreme range of possibilities for the
mean properties of galactic chemical evolution as well as heterogeneity
we show that the consistency interval $\eta \approx (2-8)\times 10^{-10}$
based upon D and $^3$He
is robust.  This strengthens the case for the nucleosynthesis determination
of the baryon density as well as the limit to the number
of light neutrino species.

\section{Stochastic Histories}

The study of the evolution of the light-element abundances within the
Galaxy has a long history (see e.g., Reeves et al.~1973; Audouze \&
Tinsley 1974).  It is a difficult problem.
Even at a given age, metal abundances in different places vary significantly.
The light-element
abundances are no different:  The D abundance measured in the nearby ISM
along different lines of sight varies significantly (Linsky 1995), and
the $^3$He abundance measured in different HII regions in the Galaxy
varies by almost an order of magnitude (Bania, Rood and Wilson 1994).
Since there is strong evidence that
chemical evolution in different parts of the Galaxy has
proceeded differently models for the mean chemical evolution cannot
be trusted to accurately represent the history in a specific location.

Our new approach allows for heterogeneity
in a most fundamental way:  we follow
the history of the material in the pre-solar
nebula through stars back to its primeval beginning.  We use
a stochastic algorithm for generating histories; from each history
the primeval D and $^3$He abundances can be determined from
pre-solar abundances.  Taking an ensemble of
histories, we construct ``a fuzzy map''
from local D and $^3$He abundances to primeval D and $^3$He abundances.

Histories are generated by a diagrammatic technique
and set of rules (see Fig.~1).  We
suppose that the pre-solar material came from the primeval mix (fraction
$f_P$) and from $N$ other stars (fractions $f_i$, $i=1,\cdots, N$).  The
fraction $f_P$ is drawn from a linear distribution whose mean is $1-\epsilon$
($\epsilon \sim 0.5$).  The number of ``first-tier stars'' $N$ is drawn from a
flat distribution whose mean is $N_0 \sim 10$; if $N< 1$, there is no material
{}from other stars and $f_P$ is set equal to one.  The fractions $f_i$ are
drawn
{}from a flat distribution whose mean is $(1-f_P)/N$.

First-tier stars are made from primeval material
and from material processed by ``second-tier stars.''
Second-tier stars are made from
primeval material and from material processed by ``third-tier stars,''
and so on.  A branch terminates
when a star is made only of primeval material.
The material from which any star is made is a fraction $f_P$
primeval and fractions $f_i$ from $N$ other stars.
At the $n$th tier, the expectation for $f_P$ is $1-\epsilon_n$
and the number of stars $N$ is drawn from a flat distribution whose
mean is $N_0\epsilon_n/\epsilon$.

A star is assumed to do the following:  (i) burn all its D to $^3$He;
(ii) return a fraction $g_3$ of its $^3$He to the ISM; and (iii)
possibly add some $^3$He and heavy elements to the material
it returns to the ISM.  The amount of $^3$He returned to the ISM by
a star is related to the D and $^3$He from which it is made
\begin{eqnarray}
 \left({{\rm D} \over {\rm H}}\right)_{\rm IN}
 & = & f_P \left({{\rm D}\over {\rm H}}\right)_P;  \\
 \left({^3{\rm He}\over {\rm H}}\right)_{\rm IN}
 & = & f_P \left( {^3{\rm He}\over {\rm H}} \right)_P + \sum_i f_i \left(
{^3{\rm He}\over {\rm H}}\right)_{\rm OUT};\\
 \left( {^3{\rm He} \over {\rm H}}\right)_{\rm OUT}
& = &  g_3 \left[ \left({{\rm D}\over {\rm H}}\right)_{\rm IN}
 +  \left({^3{\rm He}\over {\rm H}}\right)_{\rm IN} \right] + h_3 .
\end{eqnarray}
The quantities $g_3$ and $h_3$ are chosen from distributions
that are adjusted to reflect the mix of stars and our knowledge
about their processing of $^3$He.

We use oxygen as a surrogate for the heavy elements.
Massive stars produce oxygen quantified by the mass
fraction $h_{16}\sim 0.10$ of the material they return to the
ISM; low-mass stars preserve oxygen.  We require that the
oxygen mass fraction in the pre-solar material is between
0.5\% and 2\%.  The oxygen constraint
ensures that some material in the pre-solar nebula has been processed
through massive stars, but not too much.

For a given history two linear equations relate the primeval and
pre-solar D and $^3$He abundances.  The primeval D abundance is
$1/f_P$ times the pre-solar abundance.  The equation for the
primeval $^3$He abundance is more complicated, but straightforward
to obtain.  These two equations uniquely map pre-solar abundances
to big-bang abundances.   The primeval $^3$He abundance
can turn out to be negative, in which case the history must
be discarded.  This occurs when the primeval D abundance is
large (i.e., small $f_P$) and illustrates the crux
of the D + $^3$He argument:
if the primeval D abundance is large, it should lead to
a large $^3$He abundance today, and thus a negative primordial
$^3$He abundance may be required to account for the relatively
modest pre-solar $^3$He abundance.

Both the inherent uncertainty that arises
{}from not knowing the precise history of the material in the pre-solar nebula
(i.e., heterogeneity) and the uncertainty in the pre-solar abundances
themselves are treated
by Monte Carlo.  Pre-solar D and $^3$He abundances are drawn from
distributions (see below); histories are constructed as described above.

The science in our approach comes in choosing
the parameters $\epsilon$, $N_0$ and the distributions $g_3$,
$h_3$ and $h_{16}$ to reflect our understanding of galactic
chemical evolution.  The parameter $\epsilon$ controls the
fraction of material that has undergone stellar processing;
conventional wisdom has it that about 50\% of the pre-solar material
has undergone stellar processing.  The parameter $N_0$
controls the number of stars that contribute to the material
{}from which a given star is made; we have tried values from 5 to 15.

Of more importance are the distributions chosen for $g_3$, $h_3$ and $h_{16}$.
The distribution $f(g_3)$ determines the amount of $^3$He that survives
stellar processing; it in turn depends upon the stellar mass function
and the rate of return of material from stars of a given mass
to the ISM.  We parameterize $f(g_3)$
by a minimum value, $g_{\rm 3min} \sim 0.15$, and a power-law index $m$,
$f(g_3) \propto g_3^m$ for $1\ge g_3\ge g_{\rm 3min}$.  A standard mass
function and conventional stellar models correspond roughly to $m=0$
(Truran~1995).  The distribution $f(h_3;g_3)$ determines the amount of
stellar $^3$He production.  It is parameterized by $g_{3*}$:  only stars with
$g_3\ge g_{3*}\sim 0.8$ are assumed to produce $^3$He, and the
amount of $^3$He production, $h_3$, which is chosen from a flat
distribution with $0.5\times 10^{-5} \le h_3 \le 2\times 10^{-5}$.

Lastly, the distribution $f(h_{16};g_3)$ quantifies heavy-element
production by massive stars.  The distribution is characterized by
$g_{\rm 3max}\sim 0.3$, only stars with $g_3\le g_{\rm 3max}$ are
assumed to produce $^{16}$O, and $h_{16} = 0.025-0.20$, the mass
fraction of oxygen produced by massive stars which is returned to
the ISM.  The range for these two parameters is based upon
models for the yields of type II supernovae (Timmes, Woosley, \& Weaver~1995).

By varying the parameters and distributions we can explore different
models of chemical evolution, as opposed to different histories
within a model.  In this {\it Letter} we explore three models designed
to span an extreme range of possibilities.
Model 0 is chosen to be the plain, vanilla model for
chemical evolution; it is characterized by
$g_{\rm 3min} =0.1$, $g_{3*} = 0.8$, $g_{\rm 3max}=0.3$,
$\epsilon_n = \epsilon^n $, $\epsilon =0.5$, $N_0 = 10$, $h_{16} = 0.10$,
and $m=0$ for the first tier, $m=-1$ for the second tier and
so on (corresponding to more massive stars contributing to
the ISM in earlier stellar generations).
Model 1 has extreme stellar processing and $^3$He destruction;
it is characterized by $g_{\rm 3min} =0.15$,
$g_{3*}=1$ (no stellar $^3$He production), $g_{\rm 3max}=0.3$, $\epsilon_n
= \epsilon^{2n-1}$,
$\epsilon =0.8$, $N_0 = 5$, $h_{16} = 0.025$ (corresponding to
very little heavy-element return to the ISM, e.g., heavy elements
ejected with high velocity), and $m=-2$ for the
first tier, $m=-4$ for the second tier and so on.  Finally, Model 2
has less stellar processing by massive stars, more
primeval material (e.g., due to infall), and more stellar
$^3$He survival/production; it is
characterized by $g_{\rm 3min} =0.2$, $g_{3*}=0.8$, $g_{\rm 3max}=0.3$,
$\epsilon =0.25$, $\epsilon_n =\epsilon^{2n-1}$, $N_0 = 5$,
$h_{16} = 0.20$, and $m=2$ for the first tier, $m=0$ for the second tier and so
on.

These three models certainly do not
exhaust the full range of possibilities for galactic chemical evolution,
and we have studied a number of other possibilities.   An example
that illustrates the richness embodied in our approach
is a variant of Model 1 which included
very significant $^3$He production by low-mass stars.  Rather than shifting
to lower primeval abundances, D and $^3$He were
shifted to higher values.  This is because histories with many
low-mass stars in their past were excluded by the relatively low
pre-solar $^3$He abundance.   In any case, we find that
Models 0, 1, and 2 serve
well to illustrate the extreme range of possibilities.

\section{Pre-solar Abundances}

The pre-solar D and $^3$He abundances are derived
{}from $^3$He/$^4$He ratios measured in
meteorites and in the solar wind.
Because essentially all primordial D has been burned to $^3$He
in the solar convective zone and the convective zone is not hot
enough to burn $^3$He, the solar-wind value of $^3$He/$^4$He
reflects the pre-solar D + $^3$He abundance.
Measurements made using foil collectors on the moon and by
the ICI instrument on the ISEE-3 satellite give $^3$He/$^4$He
ranging from $1\times
10^{-4}$ to $10\times 10^{-4}$, depending on the phase of
the solar cycle and other factors.  Geiss and Reeves~(1972)
argue that they can correct for the
variation and that the solar-wind value is ${\rm ^3 He/ ^4 He}
= (4.1 \pm 1\,{\rm stat}) \times 10^{-4}$.
This agrees with the low-temperature component
released from carbonaceous-chondrite meteorites in step-wise
heating experiments, $^3$He/$^4$He $\simeq 4.5\times 10^{-4}$,
which is believed to be solar-wind material (Black~1972; Wieler et al.~1991).
Most recently, measurements were made at high heliographic latitudes
with the SWICS instrument on the ULYSSES spacecraft; the year-long average
isotopic ratio was determined, $^3{\rm He/^4He} = (4.4\pm 0.4)
\times 10^{-4}$, and fractionation effects were searched for and
no evidence was found (Bodmer et al., 1995).  Based upon this
measurement we adopt $^3{\rm He/^4He} = (4.4\pm 0.4)\times 10^{-4}$;
because some fractionation in the solar-wind cannot
be excluded we assign a systematic error of $1\times 10^{-4}$.

The pre-solar $^3$He/$^4$He ratio is measured in meteorites.
Black~(1972) proposed that it is
the high-temperature component released in step-wise heating of carbonaceous
chondrites (also see Eberhardt~1978), $(^3$He/$^4 {\rm He})_{\rm hi\,T}
\simeq 1.5 \times 10^{-4}$.  However, Wieler et al.\ have argued that the
high-temperature component is dominated by gas trapped in pre-solar grains
(diamonds) formed in locations far removed from the solar system and
propose that the component known as ``Q'' is a better
candidate.  Fortunately, the difference between the high-temperature
and Q components is small, $(^3{\rm He/^4 He})_Q = (1.6 \pm 0.04) \times
10^{-4}$.  A systematic error arises since no component has been
unequivocally shown to represent pre-solar $^3$He.  Taking
$(^3$He/$^4{\rm He})_Q$ as the pre-solar value, but allowing for
a systematic error to include the range encompassed by
all would-be pre-solar $^3$He values yields
$(^3$He/$^4{\rm He})_\odot = (1.6 \pm 0.04 \pm 0.3) \times 10^{-4}$.

In order to convert to abundance relative to hydrogen
one needs $({\rm ^4 He/H})_\odot$.
We use $(^4{\rm He}/{\rm H})_\odot = 0.095 \pm 0.01$, based upon
$Y_\odot = 0.27\pm 0.01$ and $Z_\odot =0.015-0.02$ as derived from
standard solar models (Turck-Chieze et al.~1988;
Bahcall \& Pinsoneault~1992).  From this we infer
\begin{eqnarray}
\left({{\rm D + ^3He \over H}}\right)_\odot & = &
  (4.2 \pm 0.7 \pm 1) \times 10^{-5}; \\
\left( {{\rm ^3 He \over H}}\right)_\odot & = &
  (1.5 \pm 0.2 \pm 0.3)\times 10^{-5}.
\end{eqnarray}

It is reassuring that the pre-solar D abundance, given by the
difference of these two numbers, is in agreement
with the HD/H$_2$ ratio measured in Jupiter, HD/H$_2 \sim
(1-3)\times 10^{-5}$ (Smith, Schempp, \& Baines~1989).
Although planetary D/H ratios are subject to chemical
fractionation, it is minimized in Jupiter since the bulk of the
deuterium exists as HD (molecular-line blanketing
does still leads to significant systematic uncertainties).

\section{Discussion}

By Monte Carlo we constructed around 300,000 histories for each
of our three chemical-evolution models.  For each history,
we draw pre-solar $^3$He and D + $^3$He abundances from a distribution
with a gaussian statistical error and top-hat systematic uncertainty.
About half of the histories are acceptable:  satisfy the
oxygen constraint (pre-solar mass fraction between 0.5\% and 2\%)
and have positive primeval $^3$He abundance.  In addition,
to ensure that the primeval D abundance is large enough to account for
that in the ISM today we weight
each point with the probability that the primordial
D abundance is greater than $(1.6\pm 0.1)\times 10^{-5}$.
Histograms of the fraction of pre-solar material that is primeval and has
been processed through 1 and 2 generations of stars as well as
the average $^3$He survival fraction are shown in Fig.~2.
The distribution of inferred
primeval D and $^3$He abundances are shown in Fig.~3;
the spread in abundances due to histories alone is also shown
and is seen to be very significant.  From
these distributions and the predicted big-bang abundances (see
Copi, Schramm, and Turner 1995) likelihood functions for the
baryon-to-photon ratio are generated (see Fig.~4).

For Models 0 and 2 (standard and high-infall model) the 95\%
confidence intervals are $\eta_{95\%} = (3.7 - 6.6)
\times 10^{-10}$ and $\eta_{95\%} = (4.1 - 7.3)\times 10^{-10}$
respectively.  For Model 1 (extreme $^3$He destruction)
the interval is shifted to lower values of the baryon-to-photon
ratio, $\eta_{95\%}= (2.1 - 4.7)\times 10^{-10}$.  This lower
bound is slightly smaller than previous ones, because we
have allowed for extreme $^3$He destruction.  In the many other models
for chemical evolution we have explored, the likelihood
function always drops precipitously
at a value of $\eta$ no smaller than $2\times 10^{-10}$---some
$^3$He necessarily survives.  While the D abundance measured in the local
ISM leads to the upper limit, $\eta \la 9\times 10^{-10}$,
consideration of the pre-solar
D abundance improves this upper bound slightly since the pre-solar D abundance
is larger.  From all this we conclude that there is a
robust concordance interval for D and $^3$He, $\eta
\approx (2- 8)\times 10^{-10}$.

This D and $^3$He consistency interval encompasses those derived by
others based upon a variety of chemical evolution models
(see e.g., Hata et al., 1994; Olive, 1995; Casse and Vangioni-Flam 1995).
Our results strongly suggest that the ``generic,'' mean chemical evolution
model of Hata et al. (1994), which is supposed to encompass
the full range of possibilities for the chemical evolution
of D and $^3$He, is less generic than the authors claim:
their 95\% confidence interval corresponds to our Model 0.

A determination of the primeval D abundance
by measuring D-Ly$\alpha$ absorption by high-redshift hydrogen clouds
could both shed light on the chemical
evolution of D and $^3$He as well as accurately determine
the baryon density.  At the moment, there are conflicting measurements
and upper limits,
and the situation is very much unsettled (York et al.~1984; Carswell et
al.~1994; Songaila et al.~1994;
Tytler et al.~1995; Rugers et al. 1995).  However, it seems likely that a
definitive determination of the primeval D abundance will be made.

\paragraph{Acknowledgments.}
We thank
Jim Truran, Frank Timmes, and Keith Olive for useful conversations about
galactic chemical evolution and Roy Lewis for his help in
understanding pre-solar D and $^3$He abundances.
This work was supported in part by the DOE (at Chicago and
Fermilab) and NASA (at Fermilab through grant NAG 5-2788 and at Chicago
through a GSRP (CJC)).

\beginapjbib

\bibitem Audouze, J. \& Tinsley, B. M.~1974, ApJ, 192, 487.

\bibitem Bahcall, J. N. \& Pinsonneault, M. H.~1992, Rev. Mod. Phys. 64, 855.

\bibitem Bania, T., Rood, R., \& Wilson, T. 1994, Proc. of the 1994
ESO Meeting on Light-element Abundances (Springer-Verlag).

\bibitem Black, D. C.~1972, Geochim. Cosmochim. Acta, 36, 347.

\bibitem Bodmer, R., Bochsler, P., Geiss, J., von Steiger,
R., \& Gloeckler, G.~1995, SpSciRev, 72, 61.

\bibitem Carswell, R. F., Rauch, M., Weymann, R. J., Cooke, A. J., \& Webb, J.
K.~1994, MNRAS, 268, L1.

\bibitem Cass\'e, M. \& Vangioni-Flam, E.~1995, The Light Element Abundances,
ed.~P.~Crane (Springer-Verlag) 44.

\bibitem Copi, C. J., Schramm, D. N., \& Turner, M. S.~1995, Science, 267, 192.

\bibitem Dearborn, D. S. P., Schramm, D. N., \& Steigman, G.~1986, ApJ, 302,
35.

\bibitem Eberhardt, P.~1978, in Proc. 9th Lunar Planet. Sci. Conf., 1027.

\bibitem Epstein, R. I., Lattimer, J. M., \& Schramm, D. N.~1976, Nature, 263,
198.

\bibitem Geiss, J.~1993, Origin and Evolution of the Elements, ed.
N.~Prantzos, E.~Vangioni-Flam, \& M.~Cass\'e (Cambridge Univ. Press) 89.

\bibitem Geiss, J. \& Reeves, H.~1972, A\&A, 18, 126.

\bibitem Hata, N., Scherrer, R. J., Steigman, G., Thomas, D., \& Walker, T.
P.~1994, astro-ph/9412087, preprint.

\bibitem Iben, I. \& Truran, J. W.~1978, ApJ, 220, 980.

\bibitem Krauss, L. M. \& Kernan, P. J.~1995, Phys. Lett. B, 347, 347.

\bibitem Linsky, J. L., etal.~1993, ApJ, 402, 694.

\bibitem Linsky, J. L.~1995, presented at the Aspen Center for
Physics Workshop on Big-bang Nucleosynthesis.

\bibitem Olive, K. A.~1995, The Light Element Abundances, ed.~P.~Crane
(Springer-Verlag) 40.


\bibitem Reeves, H. et al.~1973, ApJ, 179, 909.

\bibitem Rugers, M. et al. 1995, ApJ, submitted.

\bibitem Smith, W. H., Schempp, W. V., \& Baines, K. H.~1989, ApJ, 336, 967.

\bibitem Songaila, A., Cowie, L. L., Hogan, C. J., \& Rugers, M.~1994, Nature,
368, 599.

\bibitem Timmes, F. X., Woosley, S. E., \& Weaver, T. A.~1995, ApJS, 98, 617.

\bibitem Truran, J. W.~1995, private communication.

\bibitem Tytler, D. et al.~1995, in preparation.

\bibitem Turck-Chi\`eze, S., Cahen, S., Cass\'e, M., \& Doom, C.~1988, ApJ,
335, 415.

\bibitem Walker, T. P., Steigman, G., Schramm, D. N., Olive, K. A., \& Kang,
H.~1991, ApJ, 376, 51.

\bibitem Wieler, R., Anders, E., Baur, H., Lewis, R. S., \& Signer, P.~1991,
Geochim. Cosmochim. Acta, 55, 1709.

\bibitem Yang, J., Turner, M. S., Steigman, G., Schramm, D. N., \& Olive, K.
A.~1984, ApJ, 281, 493.

\bibitem York, D. et al.~1984, ApJ, 276, 92.

\endapjbib


\section*{Figure Captions}

\bigskip
\noindent{\bf Figure 1:}  A typical history; boxes represent stars.

\medskip
\noindent{\bf Figure 2:}  Fraction of solar-system material
that has been processed through 0, 1, and 2 stars
and the mean survival fraction of $^3$He for
Models 0 (solid), 1 (dotted) and 2 (long dash).

\medskip
\noindent{\bf Figure 3:}  Scatter plot of primeval D and
$^3$He abundances for Model 0.  The band represents the big-bang
track (prediction including $2\sigma$ theoretical uncertainty);
histograms show the distributions of predicted primeval D and $^3$He
abundances; dashed histograms illustrate the scatter
that arises from heterogeneity alone (the central values of
the D and D + $^3$He abundances were used).

\medskip
\noindent{\bf Figure 4:} Likelihood functions for $\eta$
based upon D and $^3$He abundances for models 0 (solid),
1 (dotted) and 2 (long dash).

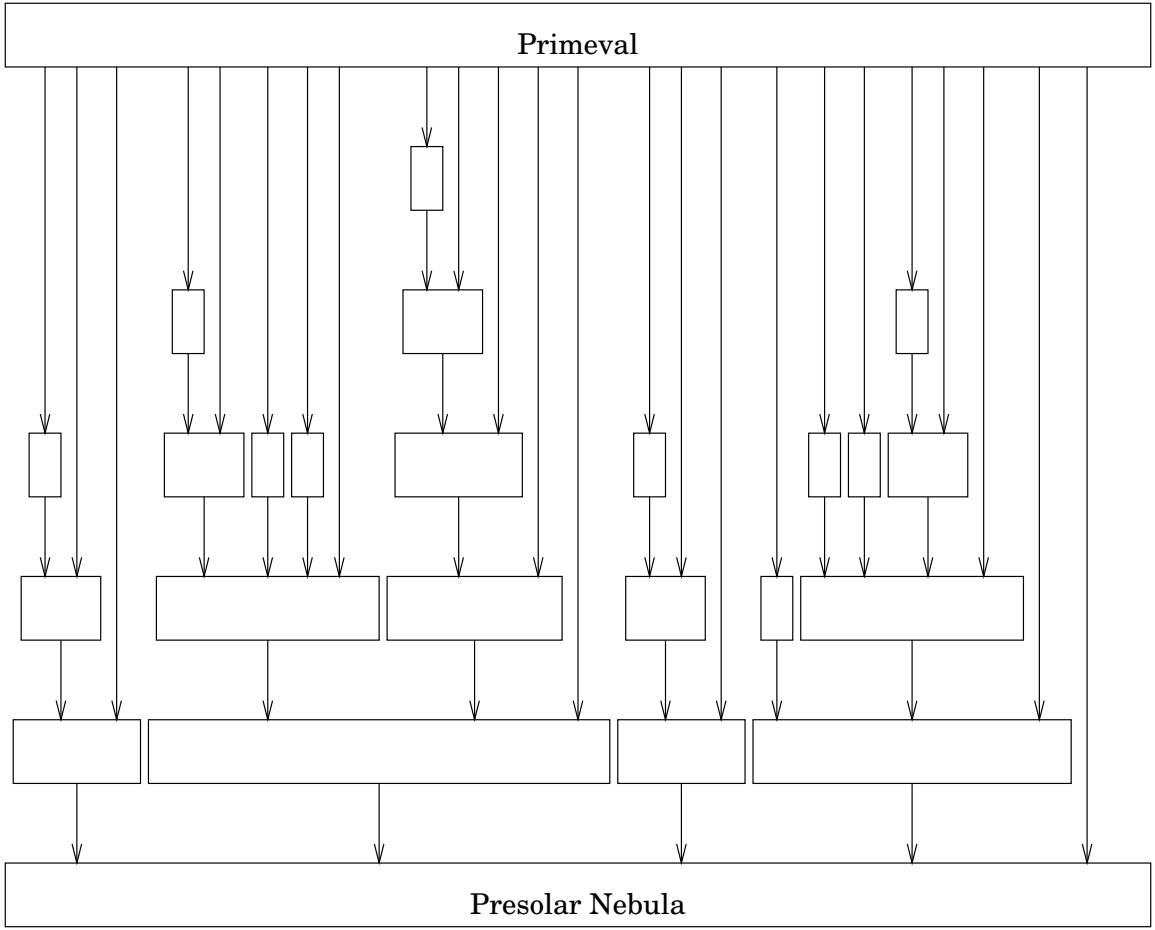
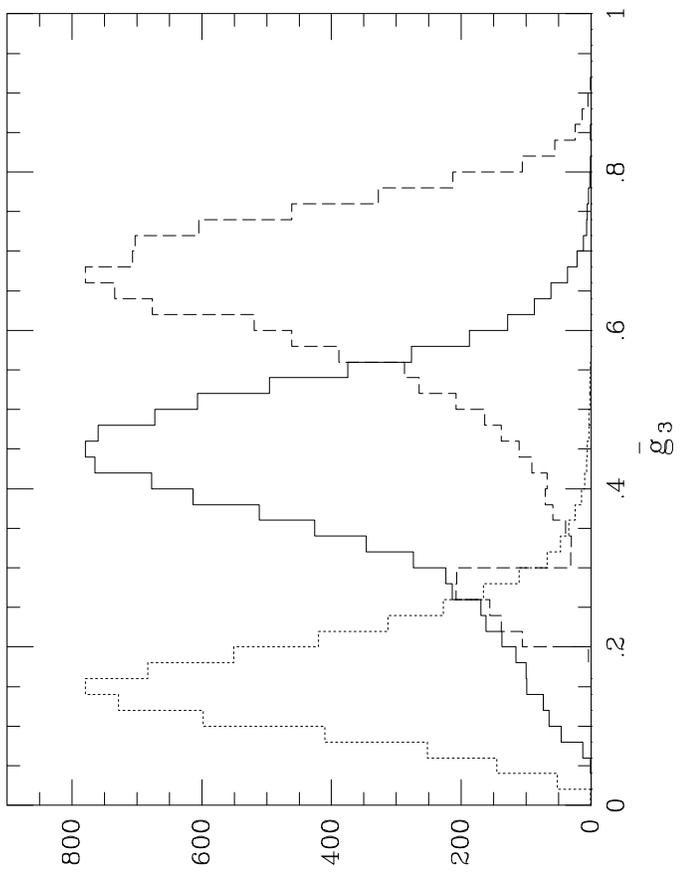
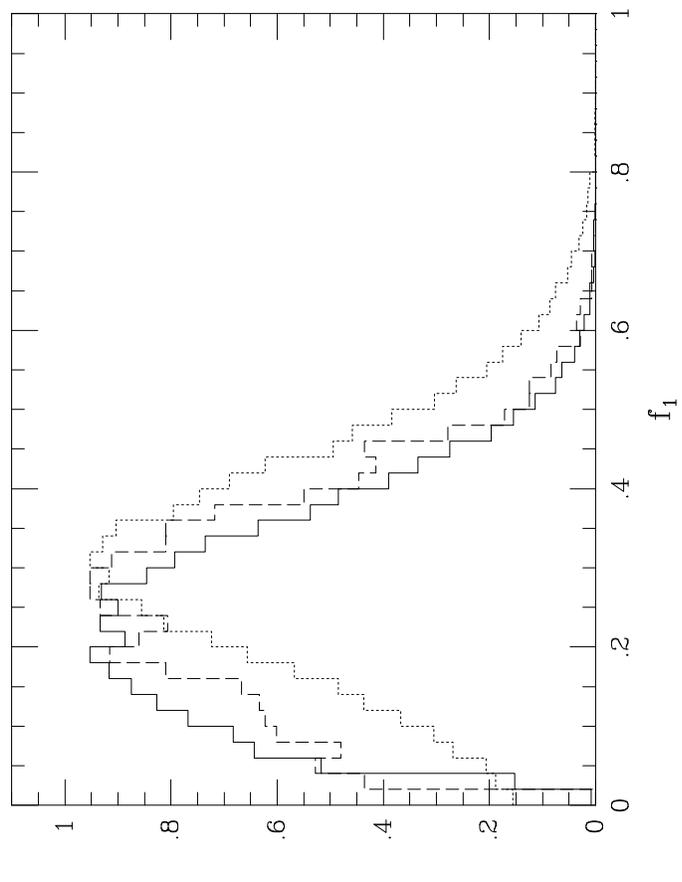
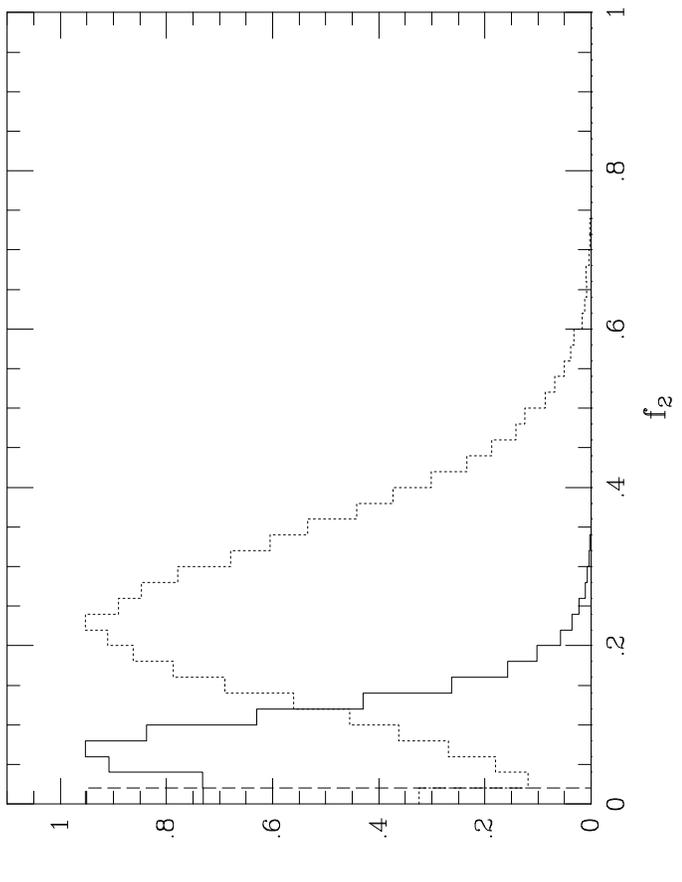
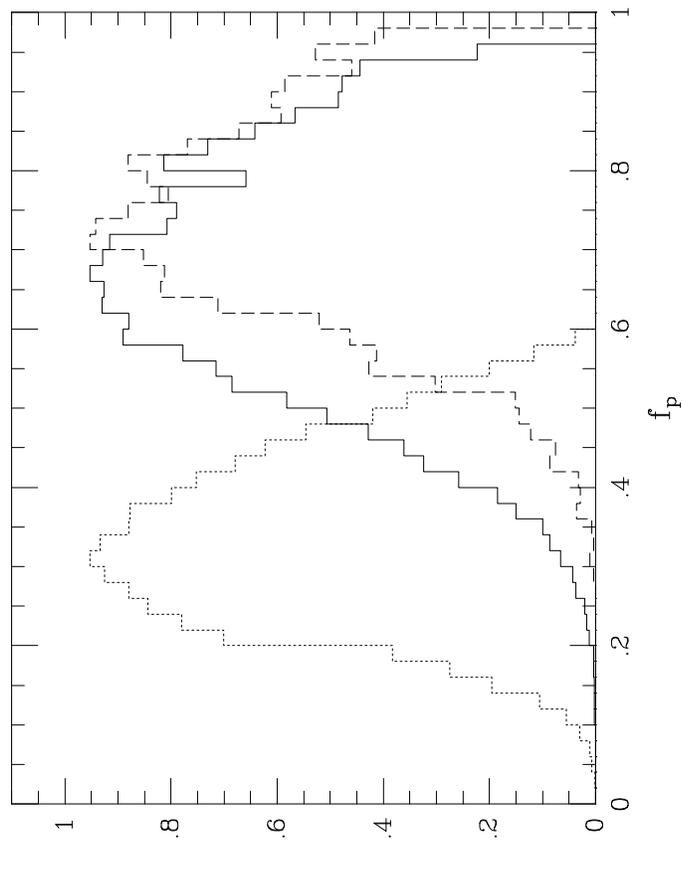
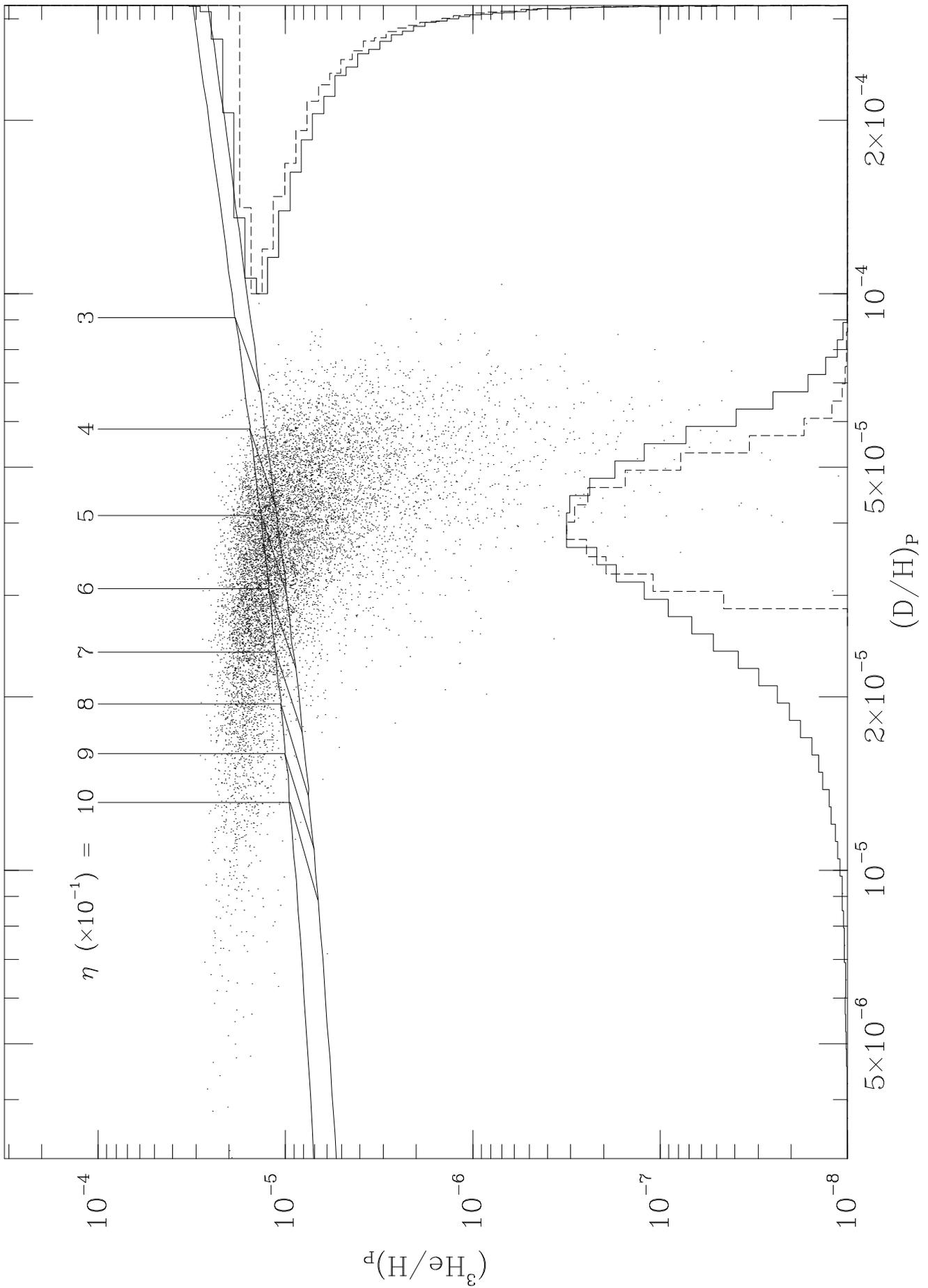
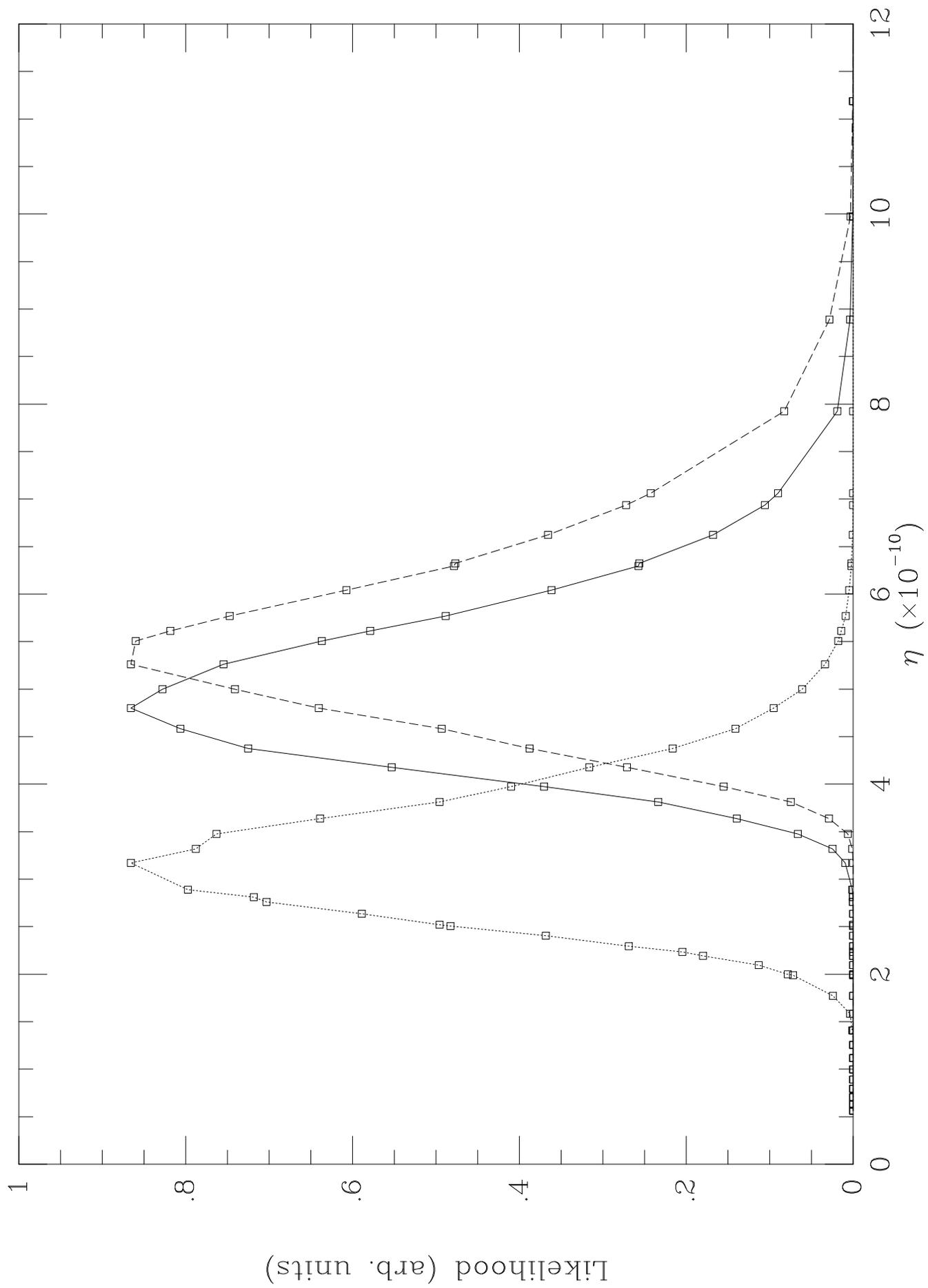
\end{document}